\documentclass[showpacs,preprintnumbers,amsmath,amssymb]{revtex4}

\usepackage{graphicx}
\usepackage{dcolumn}
\usepackage{bm}

\begin{document}


\title{Generation and Propagation of Nonlinear Waves in
Travelling Wave Tubes}

\author{Stephan I. Tzenov}
 \email{tzenov@sa.infn.it}

\affiliation{%
Dipartimento di Fisica "E.R. Caianiello", Universit\'a degli Studi
di Salerno and INFN Sezione di Napoli
\\Gruppo Collegato di Salerno, Via S. Allende, I-84081 Baronissi (SA), Italy}%

\date{\today}

\begin{abstract}
The generation and evolution of nonlinear waves in microwave
amplifiers such as travelling wave tubes, free electron lasers and
klystrons have been studied. The analysis is based on the
hydrodynamic and field equations for the self-consistent evolution
of the beam density distribution, the current velocity and the
electromagnetic fields. A system of coupled nonlinear
Schr\"{o}dinger equations for the slowly varying amplitudes of
interacting beam-density waves has been derived. Under the
approximation of an isolated mode neglecting the effect of the
rest of the modes, this system reduces to a single nonlinear
Schr\"{o}dinger equation for that particular mode.
\end{abstract}


\maketitle

\vspace{2. mm} KEY WORDS:   Microwave Amplifier, Renormalization
Group, Solitary Waves.


\vspace{0.5 cm}

\twocolumngrid

\section{\label{Intro}Introduction}

The generation and evolution of nonlinear waves and harmonic
distortions in microwave amplifiers such as travelling wave tubes,
free electron lasers (FELs) and klystrons have recently attracted
much research interest \cite{Booske,Freund,Bonifacio}. In
connection with the construction and commissioning of the next
generation of FELs and powerful klystrons for accelerating RF
cavities in circular machines and linear colliders, this issue has
become even more challenging. Of particular importance are the
effects of intense self-fields due to space charge and current, as
well as wake fields due to interaction impedances. Both of the
above influence the propagation of the electron beam in microwave
devices, its stability and transport properties. In general, a
complete description of collective processes in intense charged
particle beams is provided by the Vlasov-Maxwell equations for the
self-consistent evolution of the beam distribution function and
the electromagnetic fields. Usually, the electron beam in a
travelling wave tube can be assumed to be weakly collisional.
Hence, the dynamics of electrons is well described by the
hydrodynamic equations coupled with the equations for the
electromagnetic self-fields, which constitutes a substantial
simplification of the model. Although the analytical basis for
modelling the dynamics and behaviour of space-charge-dominated
beams is well established, a thorough and satisfactory
understanding of collective processes, detailed equilibrium and
formation of patterns and coherent structures is far from being
complete.

While the linear theory of wave generation in microwave amplifiers
is relatively well understood \cite{Dobson}, the nonlinear regime
is far from being exhaustively studied. The present paper is aimed
at filling this gap. We will be mainly interested in describing
the slow evolution of some coarse-grained quantities that are
easily measurable, such as the wave amplitudes. Owing to the
nonlinear wave interaction contingent on the nonlinear coupling
between the hydrodynamic and Maxwell equations, one can expect a
formation of nontrivial coherent structure that might be fairly
stable in space and time \cite{Tzenov,Tzenov1}. Here, we show that
solitary wave patterns in the electron beam density distribution
are an irrevocable feature, characteristic of powerful microwave
amplifiers.

The paper is organized as follows. In the next section, we state
the basic equations which will be the subject of the
renormalization group (RG) reduction in section III. Starting from
a single equation [see equation (\ref{Waveeqallord})] for the
density distribution of the electron beam, we obtain a formal
perturbation expansion of its solution to second order. As
expected, it contains secular terms proportional to powers of the
time variable which is the only renormalization parameter adopted
in our approach. In section IV, the arbitrary constant amplitudes
of the perturbation expansion are renormalized such as to
eliminate the secular terms. As a result, a set of equations for
the renormalized slowly varying amplitudes is obtained, known as
the renormalization group equations (RGEs). These equations
comprise an infinite system of coupled nonlinear Schr\"{o}dinger
equations. Finally, section V is dedicated to discussion and
conclusions.

\section{Formulation of the Problem and Basic Equations}

The electron beam in a travelling wave tube is assumed to be
weakly collisional. Therefore, the dynamics of electrons is well
described by the hydrodynamic equations coupled with the equations
for the electromagnetic self-fields. We start with the 1D
description of a beam of electrons propagating in an external
focusing electric field with focusing coefficient $G$, which
models the bunching of the electron beam in the longitudinal
direction. As we will see in the sequel its additional role is to
attain a stationary equilibrium by partially compensating the
space-charge defocusing. The continuity and the momentum balance
equations can be written as
\begin{equation}
{\frac {\partial \varrho} {\partial t}} + {\frac {\partial}
{\partial z}} {\left( \varrho v \right)} = 0, \label{Continuity}
\end{equation}
\begin{equation}
{\frac {\partial v} {\partial t}} + v {\frac {\partial v}
{\partial z}} = - G z - {\frac {k_B T} {m \varrho}} {\frac
{\partial \varrho} {\partial z}} - {\frac {e} {m}} {\left( {\frac
{\partial \varphi_{sc}} {\partial z}} + {\frac {\partial
\varphi_{w}} {\partial z}} \right)}, \label{Mombalance}
\end{equation}
where $\varrho$ and $v$ are the electron density and the current
velocity. Furthermore, $m$, $e$ and $T$ are the mass, the charge
and the temperature, respectively, while $k_B$ is the Boltzmann
constant. The space-charge $\varphi_{sc}$ and wave $\varphi_{w}$
potentials satisfy the Poisson and the wave equation
\begin{equation}
{\frac {\partial^2 \varphi_{sc}} {\partial z^2}} = - {\frac {e n
\varrho} {\varepsilon_0}}, \label{Poisseq}
\end{equation}
\begin{equation}
\Box \varphi_w = - {\frac {e n A Z_0} {c}} {\frac {\partial^2
\varrho} {\partial t^2}}, \qquad \Box = {\frac {\partial^2}
{\partial z^2}} - {\frac {1} {c^2}} {\frac {\partial^2} {\partial
t^2}}, \label{Waveequati}
\end{equation}
where $\Box$ denotes the well-known d'Alembert operator. In
addition, $\varepsilon_0$ is the permittivity of free space and $n
= N_e / V_t$ is the electron number density, where $N_e$ is the
total number of electrons and $V_t$ is the volume occupied by the
electron beam in the longitudinal direction. Moreover, the
electron beam cross-sectional area is represented by $A$, the
quantity $Z_0$ denotes the interaction impedance, while $c$
represents the phase velocity of a cold circuit wave.

Let us introduce the scaling
\begin{equation}
\varphi_{sc} = {\frac {en} {\varepsilon_0}} U_{sc}, \qquad
\varphi_{w} = {\frac {en} {\varepsilon_0}} W, \label{Scaling}
\end{equation}
and rewrite equations (\ref{Mombalance})--(\ref{Waveequati}) as
follows
\begin{equation}
{\frac {\partial v} {\partial t}} + v {\frac {\partial v}
{\partial z}} = - G z - {\frac {v_T^2} {\varrho}} {\frac {\partial
\varrho} {\partial z}} - \omega_p^2 {\left( {\frac {\partial
U_{sc}} {\partial z}} + {\frac {\partial W} {\partial z}}
\right)}, \label{Mombalance1}
\end{equation}
\begin{equation}
{\frac {\partial^2 U_{sc}} {\partial z^2}} = - \varrho, \qquad
\Box W = - {\cal Z} {\frac {\partial^2 \varrho} {\partial t^2}}.
\label{Poissequa}
\end{equation}
The electron plasma frequency $\omega_p$ and the thermal velocity
$v_T$ of the electron beam are expressed according to
\begin{equation}
\omega_p^2 = {\frac {e^2 n} {\varepsilon_0 m}}, \qquad v_T^2 =
{\frac {k_B T} {m}}, \label{Plasmafrequ}
\end{equation}
while
\begin{equation}
{\cal Z} = {\frac {\varepsilon_0 Z_0 A} {c}}, \label{Parameter}
\end{equation}
is a shorthand parameter introduced for later convenience. Note
that the thermal velocity $v_T$ as defined by equation
(\ref{Plasmafrequ}) can be alternatively expressed according to
the relation
\begin{equation}
v_T = \omega_p r_D, \qquad r_D^2 = {\frac {\epsilon_0 k_B T} {e^2
n_0}}, \label{Thermvel}
\end{equation}
where $r_D$ is the electron Debye radius. Equations
(\ref{Continuity}), (\ref{Mombalance1}) and (\ref{Poissequa})
possess a stationary equilibrium solution
\begin{equation}
\varrho_0  = {\frac {G} {\omega_p^2}}, \quad v_0 = 0, \quad U_0 =
- {\frac {G z^2} {2 \omega_p^2}}, \quad W_0 = 0. \label{Statsol}
\end{equation}
Therefore, we can further scale the hydrodynamic and field
variables as
\begin{equation}
\varrho = \varrho_0 + \epsilon R, \quad U_{sc} = U_0 + \epsilon U,
\quad v \rightarrow \epsilon v, \quad W \rightarrow \epsilon W,
\label{Scale}
\end{equation}
where $\epsilon$ is a formal small parameter introduced for
convenience, which will be set equal to one at the end of the
calculations. Thus, the basic equations to be used for the
subsequent analysis can be written in the form
\begin{equation}
{\frac {\partial R} {\partial t}} + \varrho_0 {\frac {\partial v}
{\partial z}} + \epsilon {\frac {\partial} {\partial z}} {\left( R
v \right)} = 0, \label{Continuit}
\end{equation}
\begin{equation}
{\frac {\partial v} {\partial t}} + \epsilon v {\frac {\partial v}
{\partial z}} = - {\frac {v_T^2} {\varrho_0 + \epsilon R}} {\frac
{\partial R} {\partial z}} - \omega_p^2 {\left( {\frac {\partial
U} {\partial z}} + {\frac {\partial W} {\partial z}} \right)},
\label{Mombalanc}
\end{equation}
\begin{equation}
{\frac {\partial^2 U} {\partial z^2}} = - R, \qquad \Box W = -
{\cal Z} {\frac {\partial^2 R} {\partial t^2}}. \label{Waveequat}
\end{equation}
Before we continue with the renormalization group reduction of the
system of equations (\ref{Continuit})--(\ref{Waveequat}) in the
next section, let us assume that the actual dependence of the
quantities $R$, $v$, $U$ and $W$ on the spatial variables is
represented by the expression
\begin{equation}
{\widehat{\Psi}} = {\widehat{\Psi}} {\left( z, \xi; t \right)},
\qquad {\widehat{\Psi}} = {\left( R, v, U, W \right)},
\label{Actualdep}
\end{equation}
where $\xi = \epsilon z$ is a slow spatial variable. Thus, the
only renormalization parameter left at our disposal is the time
$t$ which will prove extremely convenient and simplify tedious
algebra in the sequel.

\section{Renormalization Group Reduction of the Hydrodynamic
Equations}

Following the standard procedure of the renormalization group
method, we represent ${\widehat{\Psi}}$ as a perturbation
expansion
\begin{equation}
{\widehat{\Psi}} = \sum \limits_{n=0}^{\infty} \epsilon^n
{\widehat{\Psi}}_n, \label{Perturbexp}
\end{equation}
in the formal small parameter $\epsilon$. The next step consists
in expanding the system of hydrodynamic and field equations
(\ref{Continuit})-(\ref{Waveequat}) in the small parameter
$\epsilon$, and obtaining their naive perturbation solution order
by order. It is possible to simplify this system, which will turn
out extremely useful in what follows. Differentiating equation
(\ref{Continuit}) with respect to the time $t$, differentiating
equation (\ref{Mombalanc}) with respect to $z$ and using equations
(\ref{Waveequat}), we can eliminate the electric potentials. As a
result of obvious manipulations, we obtain
\begin{equation}
{\widehat{\cal L}} R = \epsilon \Box {\left[ {\frac {\varrho_0}
{2}} {\frac {\partial^2} {\partial z^2}} {\left( v^2 \right)} -
{\frac {\partial^2} {\partial t \partial z}} (R v) \right]}
\nonumber
\end{equation}
\begin{equation}
+ v_T^2 \Box {\frac {\partial^2} {\partial z^2}} {\left[ {\frac
{\varrho_0} {\epsilon}} \ln {\left( 1 + {\frac {\epsilon R}
{\varrho_0}} \right)} - R \right]}, \label{Waveeqallord1}
\end{equation}
where the linear differential operator ${\widehat{\cal L}} {\left(
\partial_z, \partial_t \right)}$ is given by the expression
\begin{equation}
{\widehat{\cal L}} {\left( \partial_z, \partial_t \right)} = \Box
{\left( {\frac {\partial^2} {\partial t^2}} - v_T^2 {\frac
{\partial^2} {\partial z^2}} + G \right)} + G {\cal Z} {\frac
{\partial^4} {\partial t^2 \partial z^2}}. \label{Operator}
\end{equation}
Taking into account the expansion of the logarithm, we notice that
the right-hand-side of equation (\ref{Waveeqallord1}) is at least
of first order in the formal parameter, so that
\begin{equation}
{\widehat{\cal L}} R = \epsilon \Box {\left[ {\frac {\varrho_0}
{2}} {\frac {\partial^2} {\partial z^2}} {\left( v^2 \right)} -
{\frac {\partial^2} {\partial t \partial z}} (R v) \right]}
\nonumber
\end{equation}
\begin{equation}
- \epsilon v_T^2 \Box {\frac {\partial^2} {\partial z^2}} {\left(
{\frac {R^2} {2 \varrho_0}} - {\frac {\epsilon R^3} {3
\varrho_0^2}} + {\frac {\epsilon^2 R^4} {4 \varrho_0^3}} - \dots
\right)}, \label{Waveeqallord}
\end{equation}
Equation (\ref{Waveeqallord}) represents the starting point for
the renormalization group reduction, the final goal of which is to
obtain a description of the relatively slow dynamics leading to
formation of patterns and coherent structures.

Let us proceed order by order. The solution to the zero-order
perturbation equation (\ref{Waveeqallord}) can be written as
\begin{equation}
R_0 {\left( z, \xi; t \right)} = \sum \limits_{k} A_{k} {\left(
\xi \right)} {\rm e}^{i \psi_{k} {\left( z; t \right)}},
\label{Zeroordera}
\end{equation}
where
\begin{equation}
\psi_{k} {\left( z; t \right)} = k z - \omega_{k} t, \label{Phase}
\end{equation}
and $A_{k}$ is an infinite set of constant complex amplitudes,
which will be the subject of the renormalization procedure in the
sequel. Here "constant" means that the amplitudes $A_{k}$ do not
depend on the fast spatial variable $z$ and on the time $t$,
however, they can depend on the slow spatial variables $\xi$. The
summation sign in equation (\ref{Zeroordera}) and throughout the
paper implies summation over the wave number $k$ in the case where
it takes discrete values, or integration in the continuous case.
From the dispersion equation
\begin{equation}
{\cal D} {\left( k; \omega_{k} \right)} = G {\cal Z} k^2
\omega_k^2 - \Box_k {\left( \omega_k^2 - k^2 v_T^2 - G \right)} =
0, \label{Disperequat}
\end{equation}
it follows that the wave frequency $\omega_{k}$ can be expressed
in terms of the wave number $k$, where the Fourier-image $\Box_k$
of the d'Alembert operator can be written according to
\begin{equation}
\Box_{k} = {\frac {\omega_{k}^2} {c^2}} - k^2. \label{Dalembert}
\end{equation}
It is important to emphasize that
\begin{equation}
\omega_{-k} = - \omega_k, \qquad A_{-k} = A_{k}^{\ast},
\label{Importnote}
\end{equation}
where the asterisk denotes complex conjugation. The latter assures
that the perturbed density distribution as defined by equation
(\ref{Zeroordera}) is a real quantity. The zero-order current
velocity $v_0 {\left( z, \xi; t \right)}$ obtained directly from
equation (\ref{Continuit}) can be written as
\begin{equation}
v_0 {\left( z, \xi; t \right)} = \sum \limits_{k} v_{k}^{(0)}
A_{k} {\left( \xi \right)} {\rm e}^{i \psi_{k} {\left( z; t
\right)}}, \qquad v_{k}^{(0)} = {\frac {\omega_{k}} {\varrho_0
k}}. \label{Zeroorderv}
\end{equation}

In first order equation (\ref{Waveeqallord}) acquires the form
\begin{equation}
{\widehat{\cal L}} R_1 + {\widehat{\cal L}}_z {\frac {\partial
R_0} {\partial \xi}} = \Box {\left[ {\frac {\varrho_0} {2}} {\frac
{\partial^2 v_0^2} {\partial z^2}} - {\frac {\partial^2} {\partial
t \partial z}} (R_0 v_0) - {\frac {v_T^2} {2 \varrho_0}} {\frac
{\partial^2 R_0^2} {\partial z^2}} \right]}, \label{Waveeqfirord}
\end{equation}
where by ${\widehat{\cal L}}_z$ we have denoted the derivative of
the operator ${\widehat{\cal L}}$ with respect to $\partial_z$. It
has now two types of solutions. The first is a secular solution
linearly dependent on the time variable in the first-order
approximation. As a rule, the highest power in the renormalization
parameter of the secular terms contained in the standard
perturbation expansion is equal to the corresponding order in the
small perturbation parameter. The second solution of equation
(\ref{Waveeqallord}) arising from the nonlinear interaction
between waves in the first order, is regular. Taking into account
the fact that the Fourier image of the operator ${\widehat{\cal
L}}_z$ is equal to $- i ( \partial {\cal D} / \partial k )$, we
can write the equation for determining of the secular part of the
solution as
\begin{equation}
{\widehat{\cal L}} R_1^{(s)} = i \sum \limits_k {\frac {\partial
{\cal D}} {\partial k}} {\frac {\partial A_{k}} {\partial \xi}}
{\rm e}^{i \psi_{k}}. \label{Seculfirord}
\end{equation}
We note further that
\begin{equation}
{\widehat{\cal L}} t {\rm e}^{i \psi_{k}} = t {\cal D} {\rm e}^{i
\psi_{k}} + i {\frac {\partial {\cal D}} {\partial \omega_k}} {\rm
e}^{i \psi_{k}} = i {\frac {\partial {\cal D}} {\partial
\omega_k}} {\rm e}^{i \psi_{k}}, \label{Identity}
\end{equation}
which is a direct consequence of the general relation
\begin{equation}
{\widehat{\cal L}} t G(t) = t {\widehat{\cal L}} G(t) +
{\widehat{\cal L}}_t G(t), \label{Genrelat}
\end{equation}
holding for a generic function $G(t)$. Here ${\widehat{\cal L}}_t$
implies differentiation with respect to $\partial_t$. To verify
equation (\ref{Genrelat}), it suffices to prove the identity by
induction in the case, where ${\widehat{\cal L}}$ is the monomial
operator $\partial_t^n$ and then take into account the Taylor
expansion of ${\widehat{\cal L}}$. With these remarks in hand, it
is straightforward to solve equation (\ref{Seculfirord}).
Combining its solution with the solution of the regular part, we
obtain
\begin{equation}
R_1 =- t \sum \limits_{k} u_{gk} {\frac {\partial A_{k}} {\partial
\xi}} {\rm e}^{i \psi_{k}} - {\frac {1} {2 \varrho_0}} \sum
\limits_{k,l} \alpha_{kl} A_k A_l {\rm e}^{i {\left( \psi_{k} +
\psi_{l} \right)}}, \label{Firstordvps}
\end{equation}
where $u_{gk}$ is the group velocity defined as
\begin{equation}
u_{gk} = {\frac {{\rm d} \omega_k} {{\rm d} k}} = - {\frac
{\partial {\cal D}} {\partial k}} {\left( {\frac {\partial {\cal
D}} {\partial \omega_k}} \right)}^{-1}. \label{Groupveloc}
\end{equation}
In explicit form, the components of the infinite matrix
$\alpha_{kl}$ are given by the expression
\begin{equation}
\alpha_{kl} = {\frac {\gamma_{kl}} {{\cal D}_{kl}}},
\label{Firstordaklx}
\end{equation}
where
\begin{equation}
\gamma_{kl} = \Box_{kl} (k+l) \nonumber
\end{equation}
\begin{equation}
\times {\left[ {\left( \omega_k + \omega_l \right)} {\left( {\frac
{\omega_k} {k}} + {\frac {\omega_l} {l}} \right)} - (k+l) {\left(
v_T^2 - {\frac {\omega_k \omega_l} {kl}} \right)} \right]},
\label{Firstordoper}
\end{equation}
\begin{equation}
\Box_{kl} = {\frac {{\left( \omega_k + \omega_l \right)}^2} {c^2}}
- (k+l)^2, \label{Firstordconst}
\end{equation}
\begin{equation}
{\cal D}_{kl} = {\cal D} {\left( k+l, \omega_k + \omega_l
\right)}. \label{Firstordconsta}
\end{equation}
Furthermore, the first-order  current velocity can be expressed as
\begin{equation}
v_{1} = \sum \limits_{k} v_k^{(1)} {\frac {\partial A_{k}}
{\partial \xi}} {\rm e}^{i \psi_{k}} - {\frac {1} {2 \varrho_0^2}}
\sum \limits_{k,l} \beta_{kl} A_k A_l {\rm e}^{i {\left( \psi_{k}
+ \psi_{l} \right)}}, \label{Firstordcvel}
\end{equation}
where
\begin{equation}
v_k^{(1)} = - u_{gk} v_k^{(0)} t - {\frac {i} {\varrho_0 k}}
{\left( u_{gk} - \varrho_0 v_k^{(0)} \right)},
\label{Firstordvcvel}
\end{equation}
\begin{equation}
\beta_{kl} = {\frac {\omega_k} {k}} + {\frac {\omega_l} {l}} +
\alpha_{kl} {\frac {\omega_k + \omega_l} {k+l}}, \qquad \beta_{k,
-k} = 0. \label{Currentvely}
\end{equation}

A couple of interesting features of the zero and first-order
perturbation solution are noteworthy to be commented at this
point. First of all, the zero-order density, current velocity (and
electric potentials) are proportional to the arbitrary complex
amplitudes $A_k$. The second terms in the expressions for the
first-order density $R_1$ and current velocity $v_1$ [see
equations (\ref{Firstordvps}) and (\ref{Firstordcvel})] imply
contribution from nonlinear interaction between waves. It will be
shown in the remainder that these terms give rise to nonlinear
terms in the renormalization group equation and describe solitary
wave behaviour of a generic mode.

\section{The Renormalization Group Equation}

Passing over to the final stage of our renormalization group
procedure, we note that particular terms in the second-order
perturbation equation (\ref{Waveeqallord}) will contribute to the
secular solution. Since we are interested in this solution to be
renormalized later, the contributing terms will be retained only.
Hence, we can write
\begin{equation}
{\widehat{\cal L}} R_2 + {\widehat{\cal L}}_z {\frac {\partial
R_1} {\partial \xi}} + {\frac {{\widehat{\cal L}}_{zz}} {2}}
{\frac {\partial^2 R_0} {\partial \xi^2}} = \Box {\left[ \varrho_0
{\frac {\partial^2 {\left( v_0 v_1 \right)}} {\partial z^2}}
\right.} \nonumber
\end{equation}
\begin{equation}
{\left. - {\frac {\partial^2} {\partial t
\partial z}} {\left( R_0 v_1 + R_1 v_0 \right)} - {\frac {v_T^2}
{\varrho_0}} {\frac {\partial^2 {\left( R_0 R_1 \right)}}
{\partial z^2}} + {\frac {v_T^2} {3 \varrho_0^2}} {\frac
{\partial^2 R_0^3} {\partial z^2}} \right]}, \label{Waveeqsecord}
\end{equation}
or in explicit form
\begin{equation}
{\widehat{\cal L}} R_2 = \sum \limits_k {\left[ u_{gk} {\left( -
it {\frac {\partial {\cal D}} {\partial k}} + {\frac {\partial^2
{\cal D}} {\partial k \partial \omega_k}} \right)} + {\frac {1}
{2}} {\frac {\partial^2 {\cal D}} {\partial k^2}} \right]} {\frac
{\partial^2 A_k} {\partial \xi^2}} {\rm e}^{i \psi_{k}} \nonumber
\end{equation}
\begin{equation}
+ {\frac {1} {\varrho_0^2}} \sum \limits_{k,m} \Gamma_{mk} {\left|
A_m \right|}^2 A_k {\rm e}^{i \psi_{k}}. \label{Explicit}
\end{equation}
The components of the infinite matrix $\Gamma_{mk}$ can be
expressed according to the relation
\begin{equation}
\Gamma_{mk} = \Box_k k^2 {\left[ \beta_{mk} {\left( {\frac
{\omega_m} {m}} + {\frac {\omega_k} {k}} \right)} \right.}
\nonumber
\end{equation}
\begin{equation}
{\left. - \alpha_{mk} {\left( v_T^2 - {\frac {\omega_m \omega_k}
{mk}} \right)} - v_T^2 \right]}. \label{Gammamatr}
\end{equation}
A standard algebra similar to the one outlined in the previous
section leads to the second order secular solution
\begin{equation}
R_2 = \sum \limits_{k} {\left( {\frac {u_{gk}^2 t^2} {2}} + {\frac
{G_k t} {2i}} \right)} {\frac {\partial^2 A_{k}} {\partial \xi^2}}
{\rm e}^{i \psi_{k}} \nonumber
\end{equation}
\begin{equation}
+ {\frac {t} {i \varrho_0^2}} \sum \limits_{k,m} \Gamma_{mk}
{\left( {\frac {\partial {\cal D}} {\partial \omega_k}}
\right)}^{-1} {\left| A_m \right|}^2 A_k {\rm e}^{i \psi_{k}},
\label{Secondordvps}
\end{equation}
where
\begin{equation}
G_k = {\left( u_{gk}^2 {\frac {\partial^2 {\cal D}} {\partial
\omega_k^2}} + 2 u_{gk} {\frac {\partial^2 {\cal D}} {\partial k
\partial \omega_k}} + {\frac {\partial^2 {\cal D}} {\partial k^2}}
\right)} {\left( {\frac {\partial {\cal D}} {\partial \omega_k}}
\right)}^{-1}. \label{Gkcoeffic}
\end{equation}
Taking into account the definition (\ref{Groupveloc}) of the group
velocity, we conclude that
\begin{equation}
{\frac {{\rm d} u_{gk}} {{\rm d} k}} = {\frac {\partial u_{gk}}
{\partial k}} + u_{gk} {\frac {\partial u_{gk}} {\partial
\omega_k}} = - G_k. \label{Gkcoeffgrvel}
\end{equation}

Following the standard procedure \cite{Tzenov,Tzenov1,Oono} of the
RG method, we finally obtain the desired RG equation
\begin{equation}
i {\frac {\partial {\widetilde{A}}_k} {\partial t}} + i u_{gk}
{\frac {\partial {\widetilde{A}}_k} {\partial z}} \nonumber
\end{equation}
\begin{equation}
= {\frac {G_k} {2}} {\frac {\partial^2 {\widetilde{A}}_k}
{\partial z^2}} + {\frac {1} {\varrho_0^2}} {\left( {\frac
{\partial {\cal D}} {\partial \omega_k}} \right)}^{-1} \sum
\limits_{m} \Gamma_{mk} {\left| {\widetilde{A}}_m \right|}^2
{\widetilde{A}}_k, \label{RGEquation}
\end{equation}
where now ${\widetilde{A}}_k$ is the renormalized complex
amplitude \cite{Tzenov}. Thus, the renormalized solution for the
density perturbation of the electron beam acquires the form
\begin{equation}
R {\left( z; t \right)} = \sum \limits_{k} {\widetilde{A}}_{k}
{\left( z; t \right)} {\rm e}^{i \psi_{k} {\left( z; t \right)}},
\label{Renormdens}
\end{equation}
For the renormalized electric field, we obtain
\begin{equation}
E {\left( z; t \right)} = {\frac {i e n} {\epsilon_0 G}} \sum
\limits_{k} {\frac {k^2 v_T^2 - \omega_k^2} {k}}
{\widetilde{A}}_{k} {\left( z; t \right)} {\rm e}^{i \psi_{k}
{\left( z; t \right)}}, \label{Renormelectf}
\end{equation}

Equations (\ref{RGEquation}) represent a system of coupled
nonlinear Schr\"{o}dinger equations for the amplitudes of
eigenmodes. Consider a particular mode with a wave number $k$. As
a first approximation, the contribution of the other modes with
$m\neq k$ can be neglected, which results in a single nonlinear
Schr\"{o}dinger equation for mode $k$. The nonlinearity in the
corresponding nonlinear Schr\"{o}dinger equation describes the
nonlinear interaction of the mode $k$ with itself.

It should be emphasized that the approach outlined in the present
paper is rather general even in the the case of more than one
renormalization parameter, where the extension should be
straightforward. Note that the unique assumption concerns the
knowledge of the dispersion properties of the linear differential
operator governing the evolution of the zero-order quantities. The
approach can be successfully applied to a wide class of problems
of particular physical interest.

\section{Discussion and conclusions}

We have studied the generation and evolution of nonlinear waves in
microwave amplifiers such as travelling wave tubes, free electron
lasers and klystrons. The analysis performed in the present paper
is based on the hydrodynamic and field equations for the
self-consistent evolution of the beam density distribution, the
current velocity and the electromagnetic fields. Using further the
RG method, a system of coupled nonlinear Schr\"{o}dinger equations
for the slowly varying amplitudes of interacting beam-density
waves has been derived. Under the approximation of an isolated
mode neglecting the effect of the rest of the modes, this system
reduces to a single nonlinear Schr\"{o}dinger equation for that
particular mode.

Since the approach pursued here is rather general, it is presumed
that it may find applications to other problems, where of
particular interest is the dynamics of slowly varying amplitudes
of patterns and coherent structures.

\begin{acknowledgments}
It is a pleasure to thank R.C. Davidson for many interesting and
useful discussions concerning the subject of the present paper.
\end{acknowledgments}



\onecolumngrid

\newpage

\twocolumngrid


\begin{thebibliography}{99}
\bibitem{Booske} J.G. W\"{o}hlbier, I. Dobson and J.H. Booske,
Phys. Rev. E, {\bf 66}, 056504 (2002).

\bibitem{Freund} H.P. Freund, S.G. Biedron and S.V. Milton,
IEEE Journal of Quantum Electronics, {\bf 36}, 275 (2000).

\bibitem{Bonifacio} R. Bonifacio, F. Casagrande and L. De Salvo Souza,
Phys. Rev. A {\bf 33}, 2836 (1986).

\bibitem{Dobson} J.G. W\"{o}hlbier, J.H. Booske and I. Dobson,
IEEE Transactions on Plasma Science, {\bf 30}, 1063 (2002).

\bibitem{Tzenov} S.I. Tzenov, {\it Contemporary Accelerator Physics}
(World Scientific, Singapore, 2004).

\bibitem{Tzenov1} S.I. Tzenov, New J. Phys., {\bf 6}, 19 (2004).

\bibitem{Oono} L.Y. Chen, N. Goldenfeld and Y. Oono,
Phys. Rev., {\bf E 54}, 376 (1996).

\end{thebibliography}
\end{document}